\documentclass[12pt]{article}
\pdfoutput=1 
\usepackage{amsfonts,amssymb,amsmath}

\usepackage{graphicx}
\usepackage{hyperref}
\usepackage[english]{babel}

\usepackage{natbib}

\usepackage{xspace}
\usepackage{setspace}
\usepackage{hyperref}

\usepackage[usenames]{color}         

\newcommand{\ac}{\color{red}\em }   
\newcommand{\ca}{\rm \color{black}} 
\newcommand{\cutac}[1]{\ac [cut] \ca}            

\newcommand{\cutmc}[1]{\fr [cut] \rf}            

\newcommand{\mrho}{\boldsymbol{\rho}}
\newcommand{\mmu}{\boldsymbol{\mu}}
\newcommand{\mv}{\boldsymbol{v}}
\newcommand{\mx}{\boldsymbol{x}}
\newcommand{\ms}{\boldsymbol{s}}

\begin{document}

\bibliographystyle{Chicago}

\singlespacing

\title{A Generic Multivariate Distribution for Counting Data}

\author{
Marcos Capistr\'an\thanks{Email: \textit{marcos@cimat.mx .}} ~~and J. Andr\'es Christen\thanks{ (corresponding author)
Centro de Investigaci\'on en Matem\'aticas, A. C. (CIMAT),
A.P. 402, Guanajuato, Gto. 36000, Mexico.
Tel: +52-(473)-732-7155.
Fax: +52-(473)-732-5749.
Email: \textit{jac@cimat.mx .}
}\\
Centro de Investigaci\'on en Matem\'aticas, A. C. (CIMAT)\\
Guanajuato, MEXICO.\\
}

\date{March 2011}

\maketitle

\medskip
\begin{abstract}
Motivated by the need, in some Bayesian likelihood free inference problems,
of imputing a multivariate counting distribution based on its
vector of means and variance-covariance matrix, we define a generic multivariate
discrete distribution.  Based on blending the Binomial, Poisson and  Negative-Binomial
distributions, and using a normal multivariate copula, the required distribution is
defined.  This distribution tends to the Multivariate Normal for large counts and
has an approximate pmf version that is quite simple to evaluate. 
\end{abstract}


\section{Introduction}\label{sec.intro}

We develop a generic discrete multivariate distribution defined in terms of its mean and covariate matrix only.  
The multivariate Normal distribution is defined in such terms and would be the default option as an inputed 
distributions when only the mean and covariance matrix are available for an otherwise unknown distribution.
However, there is no alternative when 
considering discrete data, specially in the case of low counts where a Normal approximation is not feasible.  
The motivation to defining such discrete distribution is as follows. 

The development and analysis of mathematical epidemic models that take into account uncertainty is an active 
field of research~\citep{breto2009time,alonso2007stochastic,finkenstadt2002stochastic,schwartz2004dynamical,nasell2002stochastic,
chen2005stochastic,andersson2000stochastic}.  The importance of this field of research is apparent given its 
potential impact on public health policies to handle emergent and re-emergent infectious diseases such as dengue 
fever, Lyme disease, tuberculosis, flu, etc. It is known that the effects of local (demographic) stochasticity 
weight more in determining the dynamics of epidemics when the number of individuals in the population is low. On 
the other hand, parameter estimation is among the standard tools to explore the predictive capacity of models 
from partial observation of the state variables. In this context, it is specially important to devise methods 
to study the predictive capacity of the mathematical models, in particular, to quantify the uncertainties.

For the sake of clarity we use the simplest epidemic model, e.g. the SIR model without vital dynamics. Let 
the random variables $X_{1}$, $X_{2}$ and $X_{3}$ denote the number of Susceptible, Infected and Recovered 
individuals in a closed population at a given time $t$, respectively. The stochastic model is defined by the processes through 
which it evolves:
\begin{align*}
 X_{1} + X_{2} &\stackrel{b_{0}\frac{X_{2}}{\Omega}}{\longrightarrow} 2X_{2}\\
 X_{2} &\stackrel{b_{1}}{\longrightarrow} X_{3}
\end{align*}
If we denote by $x_{1}$, $x_{2}$ and $x_{3}$ a realization of the random variables $X_{1}$, $X_{2}$ and $X_{3}$, 
and let $P_{\bar{x}}(t) = P(x_{1},x_{2},x_{3},t)$ be the probability that the system is in state 
$(x_{1},x_{2},x_{3})$ at time $t$, then the chemical master equation~\citep{van1992stochastic} for this system 
is given by
\begin{equation}
\label{eq:master_equation}
\begin{split}
\frac{d P(x_{1},x_{2},x_{3},t)}{dt} =& b_{0}\frac{(x_{1}+1)}{\Omega}(x_{2}-1)P(x_{1}+1,x_{2}-1,x_{3},t)\\
        &+b_{1}(x_{2}+1)P(x_{1}+1,x_{2}-1,x_{3}-1,t)\\
        &-(b_{0}\frac{x_{1}}{\Omega}x_{2} + b_{1}x_{2})P(x_{1},x_{2},x_{3},t),
\end{split}
\end{equation}
where the constant $\Omega$ denotes the total number of individuals, and $b_{0}$ and $b_{1}$ are respectively
the contact rate and the rate of loss of infectiousness.

Applying the \textit{Inverse Size Expansion}~\citep{van1992stochastic} to equation~\ref{eq:master_equation} 
leads to equations for the expected value and the fluctuations of $x_{1}$, $x_{2}$ and $x_{3}$. The Fokker-Plank 
approximation is then used leading to an approximation for the mean $E_t (X_i )$ and cross products
$E_t ( X_i X_j )$ for all species $i,j=1, 2, \ldots , k$ at each time $t$~\cite{chen2005stochastic}.  As far as 
the distribution for the $X_j$'s is concern we know to be dealing with counting data $X_i \in \mathbb{N}; i=1,2,\ldots, 
k$ and, for a fixed time $t$, we have (an approximation for) their means
$\mu_i = E_t (X_i)$, variances $v_i = E_t (X_i^2) - \mu_i^2$ and their correlations
$\rho_{ij} = \frac{E_t (X_i X_j ) - \mu_i \mu_j}{\sqrt{v_i v_j}}$.  

It is possible to simulate directly from the true model $P( \cdot ,t)$ above (for fixed $b_0$ and $b_1$) to simulate
realizations of $(X_1(t), X_2(t), X_3(t))$~\citep{gillespie2007stochastic}.  Data is commonly only available for $X_2(t)$ and for
specific epidemics (eg. Dengue fiver) there is substantial prior expert information for the contact rate, $b_0$,
and the rate of loss of infectiousness, $b_1$.
It would be possible to use the ABC algorithm~\citep{Marin2011} to make Bayesian inferences about  $b_0$ and $b_1$ but the simulation
procedure becomes very slow for moderate population sizes $\Omega$ and still the ABC approach lacks a
formal theoretical foundation~\citep[][p. 4]{Marin2011}.  Instead, using the moment approximation explained above (that indeed depends on
the parameters of interest  $b_0$ and $b_1$), we impute a counting (discrete) distribution on the
observables~\citep[commonly $X_2(t)$ but in some situations all  $X_1(t), X_2(t), X_3(t)$ are observed][]{unknown1978}
matching those moments to create a likelihood.  The computational complexity of this likelihood is in fact independent of $\Omega$ and,
using elicited priors and MCMC, a Bayesian inference is possible for any set of population sizes.  We have had
already promising results along these lines and will publish such research in an specialized journal of the field.

We will assume that the correlation matrix 
$\mrho = (\rho_{ij})$ is positively defined and, certainly, $\mu_i , v_i > 0$.  Based on this information only, 
we need to impute a discrete distribution for the observables $(X_1, X_2, X_3)$ that would be defined by these 
moments.  
Here we propose such generic multivariate distribution for counting data.  In the next section we explain the 
univariate version, which is a simple combination of the default distributions commonly used for counting data, 
namely, the Binomial, the Poisson and the Negative Binomial.  In Section~\ref{sec.multi} we create the multivariate 
version using a Normal copula and in Section~\ref{sec.exa} present some examples. 

\section{A Univariate Generic Discrete Distribution $Gd( m, v)$}\label{sec.uni}

The Poisson, Binomial and Negative Binomial distributions are simple form distributions and first candidates for 
counting data models.  For any mean and variance $\mu, v > 0$ we make a combination of these three distributions 
in the following way
\begin{equation}\label{eqn.gdd}
Gd( x | \mu, v) =
\begin{cases}
C_x^m ~ p^x (1-p)^{m-x};~ x = 0,1, \ldots m  		& \text{if}~~ \mu > v \\
e^{-v} \frac{v^x}{x!};~ x \in \mathbb{N}   				& \text{if}~~ \mu = v \\
C_{x-1}^{x+m-1} ~  p^x (1-p)^m;~  x \in \mathbb{N} 	& \text{if}~~ \mu < v \\
\end{cases} 
\end{equation}
where $p = 1 - \min \left\{ \frac{m}{v} , \frac{v}{m} \right\}$, $m = \frac{\mu^2}{| \mu - v |}$ and $C_x^m$ 
are the combinations of $m$ items taken in subsets of size $x$.  That is, we use a Binomial if  $\mu > v$, a 
Poisson if $\mu = v$ and a Negative-Binomial if $\mu < v$.  Neither of these distributions can handle 
\textit{any} mean and variance; by combining these distributions we obtain the Generic Discrete class 
$Gd( \mu, v)$ defined for arbitrary mean $\mu$ and variance $v$, $\mu,v  > 0$, and these two moments 
completely define the distribution.

Indeed, it is straightforward to see that if $X \sim Gd( \mu, v )$, $E(X) = \mu$ and $V(X) = v$.  More importantly,
for a fixed mean $\mu$, given both the properties of the Binomial and the Negative-Binomial, we see that
$$
\lim_{v \rightarrow \mu} Gd( x | \mu, v ) = e^{-v} \frac{v^x}{x!} .
$$  
Therefore we have a continuous evolution of this parametric class, being the Poisson the ``continuous bridge'' between 
the Binomial ($\mu > v$) and Negative Binomial ($\mu < v$). (Note that if $\mu > v$ and  $v \rightarrow \mu$, the 
support will increase to cover all $\mathbb{N}$ since $m \rightarrow \infty$.)

Moreover, if $X \sim Gd( \mu, v )$, $\frac{X - \mu}{\sqrt{v}}$ will tend to a standard Normal distribution if
$\mu \rightarrow \infty$ and $p \rightarrow p_0 \in (0,1)$.  That is, for large $\mu$ (and for example 
$\mu - 3\sqrt{v} > 0$) $Gd( \mu, v)$ can be approximated with a $N( \mu, v)$.  Practical guidelines for 
approximating the Poisson, Binomial and Negative-Binomial distributions should be used when calculating the cdf, 
pmf etc. of   $Gd( \mu, v)$.  We then see that the $Gd( \mu, v)$ family evolves to a Normal distribution when 
$\mu$ is large (large counts).

\section{The Multivariate Case}\label{sec.multi}

Suppose we have $k=2$ discrete distributions and we let $\mmu = ( \mu_1, \mu_2 )'$ and $\mv = ( v_1, v_2 )'$
be their vector of means and variances and $\rho = \rho_{1,2}$.  We require a bivariate discrete distribution, 
defined in terms of $\mmu$, $\mv$ and $\rho$, such that the marginal distribution for each $X_i$ is 
$Gd( \mu_i, v_i )$ and the resulting correlation between $X_1$ and $X_2$ is (at least approximately) $\rho$.

We use a Normal Copula~\citep[see][chap. 2]{Nelsen:2006} to create a a joint distribution.  Let
$$
\phi_{\rho} ( s, t ) = \frac{1}{2 \pi \sqrt{1-\rho^2}} \exp \left\{ -\frac{( s^2 - 2 \rho s t + t^2)}{2 (1 - \rho^2)} \right\}
$$ 
be the bivariate standard normal distribution with correlation $\rho$ and $\Phi(s)$ the standard normal cdf.  
The normal Copula is defined as
$$
C_{\rho} ( u, v) = \int_{-\infty}^{\Phi^{-1}(u)} \int_{-\infty}^{\Phi^{-1}(v)} \phi_{\rho} ( s, t ) ds dt .
$$
We define the joint cdf of $X_1$ and $X_2$ as
$$
F_{\mmu, \mv, \rho}( x_1, x_2 ) = C_{\rho} ( F_{\mu_1, v_1}(x_1), F_{\mu_2 , v_2}(x_2) ) ,
$$
where $F_{\mu, v}$ is the cdf of a $Gd( \mu, v )$.  $F_{\mmu, \mv, \rho}( x_1, x_2 )$ defines the Generic Discrete 
distribution in dimension 2, $Gd_2( \mmu, \mv, \rho)$, and it is straightforward to verify that its pmf is
\begin{eqnarray}\label{eqn.pmf}
Gd_2( x_1, x_2 | \mmu, \mv, \rho ) & = &
         C_{\rho} ( F_{\mu_1, v_1}(x_1), F_{\mu_2 , v_2}(x_2) ) - C_{\rho} ( F_{\mu_1, v_1}(x_1-1), F_{\mu_2 , v_2}(x_2) ) - \\
&    & C_{\rho} ( F_{\mu_1, v_1}(x_1), F_{\mu_2 , v_2}(x_2-1) ) + C_{\rho} ( F_{\mu_1, v_1}(x_1-1), F_{\mu_2 , v_2}(x_2-1) )  \nonumber
\end{eqnarray}
(that is, the pmf are the corresponding jumps in the stepped cdf $F_{\mmu, \mv, \rho}( x_1, x_2 )$).
Since $C_{\rho} ( u, v)$ is a Copula for every $-1 \leq \rho \leq 1$, the marginal distributions will be precisely 
$Gd( \mu_1, v_1 )$ and $Gd( \mu_2, v_2 )$, as required~\citep{Nelsen:2006}.  

By pretending $F_{\mu_i, v_i}(x_i)$ is differentiable, with derivative $Gd( x_i | \mu_i, v_i )$,
`differentiating' $F_{\mmu, \mv, \rho}( x_1, x_2 )$ suggest the following approximation to the pmf
\begin{equation}\label{eqn.approx}
Gd_2( x_1, x_2 | \mmu, \mv, \rho ) \approx gd_2( x_1, x_2 | \mmu, \mv, \rho ) = K \frac{\phi_{\rho} ( s, t )}{\phi( s ) \phi( t )}
Gd( x_1 | \mu_1, v_1 ) Gd( x_2 | \mu_2 , v_2 )
\end{equation}
were $s =  \Phi^{-1}(F_{\mu_1, v_1}(x_1))$, $t = \Phi^{-1}(F_{\mu_2 , v_2}(x_2))$ and $\phi(\cdot)$ is the 
standard Normal pdf, for some normalization constant $K$. This approximation will prove useful even for $\mu_i$ 
small, and is far less computationally demanding than the exact version in (\ref{eqn.pmf}).

If $F_{\mu_i, v_i}(x_i)$ were cdf's of $N( \mu_i, v_i )$, that is 
$F_{\mu_i, v_i}(x_i) = \Phi\left( \frac{x_i - \mu_i}{\sqrt{v_i}} \right)$, we immediately see that the correlation 
between $X_1$ and $X_2$ is $\rho$.  In the general case as each marginal distribution $Gd( \mu_i , v_i )$ becomes 
similar to a Normal distribution the actual correlation will be approximately $\rho$ and $Gd_2$ becomes increasingly
similar to a Bivariate Normal distribution with the correct moments.

Finally, the multivariate generic discrete distribution is defined as
$$
Gd_n( \mx | \mmu, \mv, \mrho ) = 
\int_{\Phi^{-1}(F_{\mu_1, v_1}(x_1-1))}^{\Phi^{-1}(F_{\mu_1, v_1}(x_1))} \cdots
\int_{\Phi^{-1}(F_{\mu_n , v_n}(x_n-1))}^{\Phi^{-1}(F_{\mu_n , v_n}(x_n))}
(2 \pi)^{-n/2} |\mrho|^{-1/2} \exp \left\{ -{1 \over 2} \ms' \mrho^{-1} \ms \right\} ds_1 \cdots ds_n .
$$
The corresponding approximate pmf would be
$$
gd_n( \mx | \mmu, \mv, \mrho ) = K \frac{\phi_{\mrho} ( s_1, s_2, \ldots , s_ n)}{\phi( s_1 ) \phi( s_2 ) \cdots \phi(s_n) }
Gd( x_1 | \mu_1, v_1 ) Gd( x_2 | \mu_2 , v_2 ) \cdots Gd( x_n | \mu_n, v_n ) ,
$$
with  $s_i =  \Phi^{-1}(F_{\mu_i, v_i}(x_i))$.

\section{Example}\label{sec.exa}

\begin{figure}
\begin{center}
\begin{tabular}{c c}
\includegraphics[height=7cm, width=7cm]{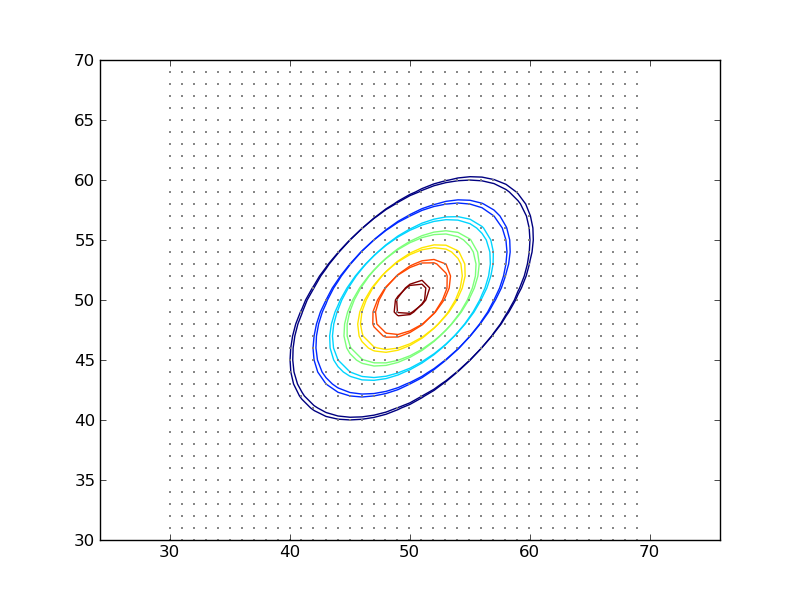} &
\includegraphics[height=7cm, width=7cm]{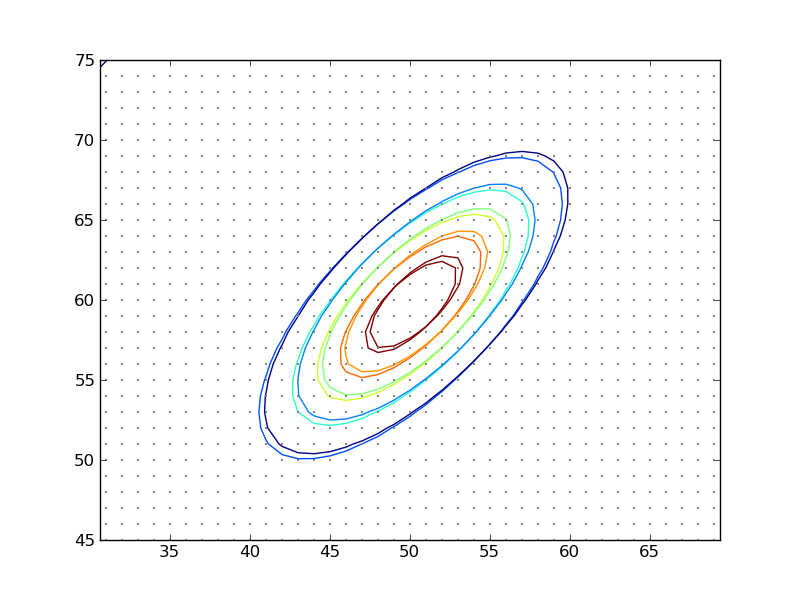} \\
(a) & (b) \\
\includegraphics[height=7cm, width=7cm]{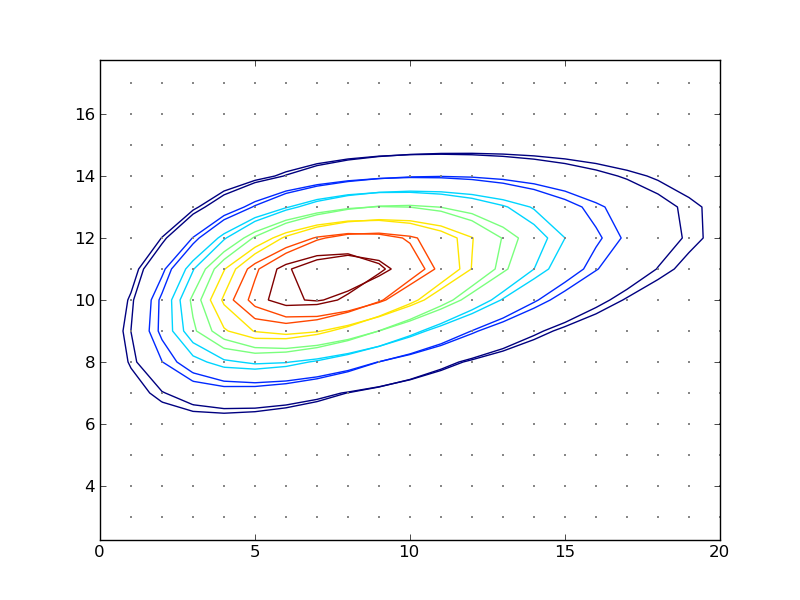} &
\includegraphics[height=7cm, width=7cm]{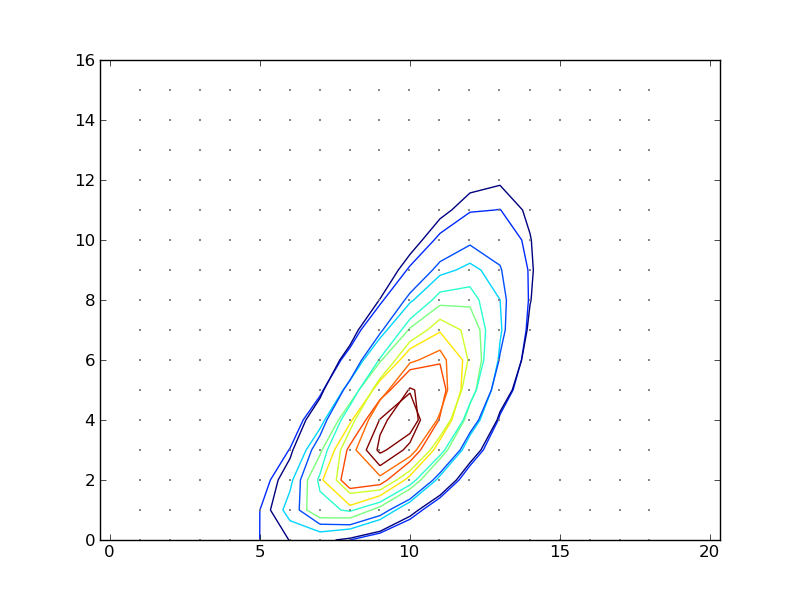} \\
(c) & (d) \\
\end{tabular}
\caption{\label{fig.exa} Contour plots  for the exact and approximate pmf's of $Gd_2$, for the parameters described
in Table~\ref{tab.exa}.  The contours in all cases basically match.}
\end{center}
\end{figure}

We present four examples of $Gd_2$, see Figure~\ref{fig.exa} and Table~\ref{tab.exa}.  We compare the 
approximation with the exact distributions by calculating numerically the exact pmf in  (\ref{eqn.pmf})
and the approximate pmf, $gd_2$, in (\ref{eqn.approx}) over a relevant grid of the support for each example.
The approximation seems to be very good option and far less computationally demanding.
Moreover, the moments match correctly and both $Gd_2$ and $gd_2$ have an actual correlation
quite near the required one (compare $\rho$ with $\rho'$ and $\rho^{*}$ in Figure~\ref{tab.exa}).
It is also very remarkable that the normalization constant needed for the approximation is quite close
to 1.  This will potentially enable the use of $gd_n$ as an alternative, less computationally demanding
likelihood, by considering $K$ to depend only marginally on the mean and variance-covariance matrix.

\begin{table}
\begin{tabular}{ c | r r r r r | r | r r r r r r }
  & $\mu_1$ & $v_1$ & $\mu_2$ & $v_2$ & $\rho$ & $\rho'$ & $K$ & $\mu_1^*$ & $v_1^*$ & $\mu_2^*$ & $v_2^*$ & $\rho^{*}$ \\ \hline
(a) & 50 & 25 & 50 & 25 &    0.5 & 0.5144 &0.99 & 50.25 & 24.99 & 50.25 & 24.99 & 0.5009 \\ 
(b) & 50 & 25 & 60 & 25 &    0.7 & 0.7129 &0.99 & 50.35 & 25.04 & 59.85 & 24.76 & 0.7013 \\ 
(c) & 10 & 25 & 11 &  4 &    0.4 & 0.3988 &0.99 &  9.45 & 23.30 & 10.90 &  3.85 & 0.3876 \\ 
(d) & 10 &  5 &  5 & 10 &    0.7 & 0.6869 &0.97 & 10.23 &  4.68 &  5.43 & 10.43 & 0.6670 \\ 
\end{tabular}
\caption{\label{tab.exa} $Gd_2$, resulting correlation, $\rho'$,  for the exact and resulting
moments for the approximate (*) pmf of $Gd_2$, $gd_2$, with various parameters.  Contour plots
for the exact and approximate pmf's of distribution (a)-(d) may be seen in Figure~\ref{fig.exa}.
The normalization constant needed for $gd_2$ is given in column $K$.}
\end{table}

\section{Discussion}

We develop a generic discrete multivariate distribution defined in terms of its vector of means
and variance-covariance matrix only, as it is the case for the Multivariate-Normal distribution for
continuous data.  This distribution has applications in the Bayesian analysis of complex models
were we are dealing with counting data and the correct likelihood is not available analytically,
but approximation techniques can be developed to obtain moments of observables.
This is the case when studying epidemics using the SIR stochastic model, as explained
in Section~\ref{sec.intro}.  The distribution developed here can now be used as a default
distribution to be imputed to multivariate counting data in such situations.  Moreover,
when large counts are involved this distribution tends to a Multivariate Normal,
(eg. Figure~\ref{fig.exa}(a) and (b)). 

\bibliography{GenDiscDist}

\end{document}